\documentclass[conference]{IEEEtran}
\IEEEoverridecommandlockouts

 \pdfoutput=1

\usepackage{cite}
\usepackage{amsmath,amssymb,amsfonts}
\usepackage{algorithmic}
\usepackage{graphicx}
\usepackage{textcomp}
\usepackage{xcolor}

\usepackage{amsmath,amssymb,amsfonts}
\usepackage{algorithmic}
\usepackage{graphicx}
\usepackage{textcomp}
\usepackage{xcolor}
\usepackage{longtable}
\usepackage{float}
\usepackage{graphicx}
\usepackage{multirow}
\usepackage{balance}
\usepackage{flushend}
\usepackage{hyperref}
\usepackage{tabularx} 
\usepackage{booktabs} 
\usepackage{hhline}
\usepackage{enumitem}
\usepackage{array}
\newcolumntype{L}[1]{>{\raggedright\let\newline\\\arraybackslash\hspace{0pt}}m{#1}}
\newcolumntype{C}[1]{>{\centering\let\newline\\\arraybackslash\hspace{0pt}}m{#1}}
\newcolumntype{R}[1]{>{\raggedleft\let\newline\\\arraybackslash\hspace{0pt}}m{#1}}
\usepackage{subfigure}

\newenvironment{boxedtext}
    {
    
    \begin{center}

    \begin{tabular}{|p{0.96\linewidth}|}
    \hline
    }
    { 
    \\ \hline
    \end{tabular} 
    
    \end{center}
       }

\def\BibTeX{{\rm B\kern-.05em{\sc i\kern-.025em b}\kern-.08em
    T\kern-.1667em\lower.7ex\hbox{E}\kern-.125emX}}
\begin{document}

\title{Why Security Defects Go Unnoticed during Code Reviews? A Case-Control Study of the Chromium OS Project}

\author{\IEEEauthorblockN{Rajshakhar Paul, Asif Kamal Turzo, Amiangshu Bosu}
\IEEEauthorblockA{\textit{Department of Computer Science} \\
\textit{Wayne State University}\\
Detroit, Michigan, USA \\
\{r.paul, asifkamal, amiangshu.bosu\}@wayne.edu }
}

\maketitle

\begin{abstract}
Peer code review has been found to be effective in identifying security vulnerabilities. However, despite practicing mandatory code reviews, many Open Source Software (OSS) projects still encounter a large number of post-release security vulnerabilities, as some security defects escape those. Therefore, a project manager  may wonder if there was any weakness or inconsistency during a code review that missed a security vulnerability. Answers to this question may help a manager pinpointing areas of concern and taking measures to  improve the effectiveness of his/her project's code reviews in identifying security defects.
Therefore, this study aims \textit{to identify the factors that differentiate code reviews that successfully identified security defects from those that  missed such defects.}

With this goal, we conduct a case-control study of Chromium OS project. Using multi-stage semi-automated approaches, we build a dataset of 516 code reviews that successfully identified security defects and 374 code reviews where security defects escaped. The results of our empirical study suggest that  the are significant differences between the categories of security defects that are identified and  that  are missed during code reviews. A logistic regression model fitted on our dataset achieved an AUC score of 0.91 and has identified nine code review attributes that  influence identifications of security defects. While time to complete a review, the number of mutual reviews between two developers, and if the review is for a bug fix have  positive impacts on vulnerability identification, opposite effects are observed from the number of directories under review, the number of total reviews by a developer, and the total number of prior commits for the file under review.
\end{abstract}


\section{Introduction}
\label{sec:intro}

Peer code review (a.k.a. code review) is a software quality assurance practice of getting a code change inspected by peers before its integration to the main codebase. In addition to improving maintainability of a project and identification of bugs~\cite{bosu2016process,bacchelli2013expectations}, code reviews have been found useful in preventing security vulnerabilities~\cite{bosu-fse14,mcgraw-security-in}. Therefore, many popular  Open Source Software (OSS) projects such as, Chromium, Android, Qt, oVirt, and Mozilla as well as commercial organizations such as, Google, Microsoft, and Facebook have integrated code reviews in their software development pipeline~\cite{bosu2016process,rigby2013convergent}. With mandatory code reviews, many OSS projects (e.g., Android, and Chromium OS) require each and every change to be reviewed and approved by multiple peers~\cite{rigby2013convergent,bacchelli2013expectations}. 
Although mandatory code reviews are preventing a significant number of security defects~\cite{munaiah2017natural,bosu-fse14}, these projects still report a large number of post-release security defects in the Common Vulnerabilities and Exposure (a.k.a. CVE) database\footnote{\url{https://cve.mitre.org/cve/}}. Therefore, a project manager from such a project may wonder if there was any weakness or inconsistency during a code review that missed a security vulnerability. For example, she/he may want to investigate:  i) if reviewers had adequate expertise relevant to a particular change, ii) if reviewers spent adequate time on the reviews, or iii) if the code change was too difficult to understand. Answers to these questions may help a manager pinpointing areas of concern and taking measures to  improve the effectiveness of his/her project's code reviews in identifying security defects.

To investigate these questions, this study aims \textit{to identify the factors that differentiate code reviews that successfully identified security defects from those that  missed such defects.} Since code reviews can identify vulnerabilities very early in the software development pipeline,  security defects identified during code reviews incur significantly less cost, as the longer it takes to detect and fix a security vulnerability, the more that vulnerability will cost~\cite{mcgraw2008automated}.  Therefore, improving the effectiveness of code reviews in identifying security defects may reduce the cost of developing a secure software.

With this goal, we conducted a case-control study of the Chromium OS project. Case-control studies, which are common in the medical field, compare two existing groups differing on an outcome~\cite{setia2016methodology}. We identified the cases and the controls based on our outcome of interest, namely
whether a security defect was identified or escaped during the  code review of a vulnerability contributing commit (VCC).  Using a keyword-based mining approach followed by manual validations on a dataset of 404,878 Chromium OS code reviews, we identified 516 code reviews that successfully identified security defects. In addition, from the Chromium OS bug repository, we identified 239 security defects that escaped code reviews. Using a modified version of the SZZ algorithm~\cite{Borg_2019} followed by manual validations, we identified 374 VCCs and corresponding code reviews that  approved those changes. Using these two datasets, we conduct an empirical study and answer the following two research questions:

\begin{enumerate}[start=1,label={(\bfseries RQ\arabic*):}]

\item \textbf{Which categories of security defects are more likely to be missed during code reviews?}

\underline{Motivation:} Since a reviewer primarily relies on his/her knowledge and understanding of the project,  some categories of security defects may be more challenging to identify during code reviews than others. The results of this investigation can help a project manager in two ways. First, it will allow a manager to leverage other testing /quality assurance methods  that are more effective in identifying categories of vulnerabilities that are more likely to be missed during code reviews. Second, a manager can arrange training materials  to  educate developers and adopt more effective code review strategies for those categories of security vulnerabilities.

\underline{Findings:} The results suggest that some categories of vulnerabilities are indeed more difficult to identify during code reviews than others. The identification of a vulnerability that requires an understanding of a few lines of the code context (e.g., unsafe method, calculation of buffer size, and resource release) are more likely to be identified during code reviews. On the other hand, vulnerabilities that require either code execution (e.g., input validation) or understanding of  larger code contexts (e.g., resource lifetime, and authentication management ) are more likely to remain unidentified.

\item \textbf{Which factors influence the identification of security defects during a code review?}

\underline{Motivation:}

Insights obtained from this investigation can help a project manager pinpoint areas of concern and take targeted measures to improve the effectiveness of his/her project's code reviews in identifying security defects.

\underline{Findings:}
We developed a Logistic Regression model based on 18 code review attributes. The model, which achieved an AUC of 0.91, found nine code review attributes that distinguish code reviews that missed a vulnerability from the ones that did not. According to the model, the likelihood of a security defect being identified during code review declines with the increase in  the number of directories/files involved in that change. Surprisingly, the likelihood of missing a vulnerability during code reviews increased with a developer's reviewing experience. Vulnerabilities introduced in a bug fixing commit were more likely to be identified than those introduced in a non-bug fix commit. 
\end{enumerate}


\textbf{The primary contributions} of this paper are:
\begin{itemize}
    \item An empirically built and validated dataset of code reviews that either identified or missed security vulnerabilities.
    \item An  empirical investigation of security defects that escaped vs. the ones that are identified during code reviews.
    \item A logistic regression model to identify relative importance of various factors influencing identification of security defects during code reviews.
    \item An illustration of conducting a case-control study in the software engineering context. 
    \item We make our script and the dataset publicly available at:  {\url{https://zenodo.org/record/4539891}}.
\end{itemize}

\textbf{Paper organization:} The remainder of this paper is organized as follows.
Section~\ref{sec:background} provides a brief background on code reviews and case-control study. 
Section~\ref{sec:method} details our research methodology.
Section~\ref{sec:results} describes the results of our case-control study.
Section~\ref{sec:implication} discusses the implications based on the results of this study. 
Section \ref{sec:threats} discusses the threats to validity of our findings.
Section~\ref{sec:related-work} describes related works. 
Finally, Section \ref{conclusion} provides the future direction of our work and concludes this paper. 

\section{Background}\label{sec:background}

This section provides a brief background on security vulnerabilities, code reviews, and case-control studies.

\subsection{Security Vulnerabilities}
A vulnerability is a weakness in a software component that can be exploited by a threat actor, such as an attacker, to perform unauthorized actions within a computer system. Vulnerabilities result mainly from bugs in code which arise due to violations of secure coding practices, lack of web security expertise, bad system design, or poor implementation quality. Hundreds of types of security vulnerabilities can occur in code, design, or system architecture. The security community uses Common Weakness Enumerations (CWE)~\cite{cwe2020} to provide an extensive catalog of those vulnerabilities. 

\subsection{Code reviews}

Compared with the traditional heavy-weight inspection process, peer code review is more light-weight, informal, tool-based, and used regularly in practice~\cite{bacchelli2013expectations}. In addition to their positive effects on software quality in general, Code review can be an important practice for detecting and fixing security bugs early in a software development lifecycle~\cite{mcgraw-security-in}.
For example, an expert reviewer can identify potentially vulnerable code and help the author  to fix the vulnerability or to abandon  the code. 
Peer code review can also identify attempts to insert malicious code.
Software development organizations, both OSS and commercial, have been increasingly adopting tools to manage the peer code review process~\cite{rigby2013convergent}. A tool-based code review starts, when an author creates a patch-set (i.e. all files added or modified in a single revision), along with a description of the changes, and submits that information to a code review tool. After reviewers are assigned, the code review tool then notifies selected reviewers regarding the incoming request. During a review, the tools may highlight the changes between revisions in a side-by-side display. The review tool also facilitates communication between the reviewers and the author in the form of review comments, which may focus on a particular code segment or the entire patchset. 
By uploading a new patchset to address the review comments, the author can  initiate a new review iteration. This review cycle repeats until either the reviewers approve the change or the author abandons.
If the reviewers approve the changes, then the author commits the patchset or asks a project committer to integrate the patchset to the project repository.

\subsection{Case-Control Study}
Case-control studies, which are widely used in the medical field, is a type of observational study, where subjects are selected based on an outcome of interest, to identify factors that may contribute to a medical condition by comparing those with the disease or outcome (cases) against a very similar group of subjects who do not have that disease or outcome  (controls)~\cite{lewallen1998epidemiology}. A case-control study is always retrospective, since it starts with an outcome  and then traces back to investigate exposures. However, it is essential that case inclusion criteria are clearly defined to ensure that all cases included in the study are based on the same diagnostic criteria.  To measure the strength of the association between a given exposure and an outcome of interest, researches who conduct case-control studies usually use Odds Ratio (OR), which represents the odds that an outcome will occur given an exposure, compared to the odds of the outcome occurring in the absence of that exposure~\cite{setia2016methodology}. 

Although case control studies are predominantly used in the medical domain, other domains have also used this research method. In the SE domain, Allodi and Massaci conducted case-control studies to investigate vulnerability severities and their exploits~\cite{allodi2014comparing}.
In a retrospective study, where two groups naturally emerge based on an outcome, the case-control study framework can provide researchers guidelines in  selecting variables, analyzing data, and reporting results.
We believe that a case-control design is appropriate for this study, since we are conducting a retrospective study, where two groups differ based on an outcome. In our design, each of the selected cases is a vulnerability contributing commit forming two groups: 1) cases--vulnerabilities identified during code reviews  and 2) controls --vulnerabilities escaped code reviews.

\section{Research Methodology}
\label{sec:method}
Our research methodology focused on identifying vulnerability contributing commits that went through code reviews and had its' security defects either getting identified or escaping. In the following subsections, we detail our research methodology.

\subsection{Project Selection}
For this study, we select the Chromium OS project for the following five reasons-- (i) it is one of the most popular OSS projects, (ii) it is a large-scale matured project containing more than 41.7 million Source Lines of Code (SLOC) \cite{openhub-chromium}. (iii) it has been conducting tool-based code reviews for almost a decade, (iv) it maintains security advisories\footnote{\url{https://www.chromium.org/chromium-os/security-advisories}} to provide regular updates on identified security vulnerabilities, and (v) it has been subject to prior studies on security vulnerabilities~\cite{meneely2013patch, camilo2015bugs, munaiah2016vulnerability, lagerstrom2017exploring, meneely2014empirical}.

\subsection{Data Mining}
The code review repositories of the  Chromium OS project is managed by Gerrit\footnote{https://www.gerritcodereview.com/} and is publicly available at: \url{https://chromium-review.googlesource.com/}. We wrote a Java application to access Gerrit's REST API to mine all the publicly available code reviews for the project and store the data in a MySQL database. Overall, we mined 404,878 code review requests spanning March 2011 to March 2019. Using an approach similar to Bosu et al.~\cite{Bosu-Carver-ESEM:2014}, we  filtered the bot accounts, using a set of keywords (e.g., `bot', `CI', `Jenkins', `build', `auto', and `travis') followed by manual validations, to exclude the comments not written by humans.
To identify whether multiple accounts belong to a single person, we follow a similar approach as Bird \textit{et al.}~\cite{bird2006mining},  where we use the Levenshtein distance between two names to identify similar names. If our manual reviews of the associated accounts suggest that those belong to the same person, we merge those to a single account.

\subsection{Building a  dataset of cases (i.e. vulnerabilities identified during code reviews)} We adopted a keyword-based semi-automated  mining approach, which is similar to the strategy used by Bosu \textit{et} al.~\cite{bosu-fse14}, to build a dataset of vulnerabilities identified during code reviews. Our keyword-based mining was based on the following three steps:

 \emph{(Step I) Database search:} We queried our MySQL database of Chromium OS code reviews to select review comments that contain at least one of the 105 security-related keywords (Table~\ref{table:security-keywords}). Bosu \textit{et} al.~\cite{bosu-fse14} empirically developed and validated a list of 52 keywords to mine code review comments associated with the 10 common types of security vulnerabilities. Using Bosu \textit{et} al's  keyword list as our starting point, we added additional 53 keywords to this list based on the NIST glossary of security terms~\cite{kissel2011glossary}.  Our database search identified 7,572 code review comments that included at least one of these 105 keywords (Table~\ref{table:security-keywords}).

 \emph{(Step II) Preliminary filtering:} Two of the authors independently audited each code review comment identified during the database search to eliminate any reviews that clearly did not raise a security concern. We excluded a review comment in this step only if both auditors independently determined that the comment does not refer to a security issue.
To illustrate the process let's examine two code review comments with the same keyword `overflow'. The first comment-- ``\textit{no check for overflow here?}" potentially raises a concern regarding an unchecked integer overflow and therefore was included for a detailed inspection. While the second comment --``\textit{I'm not sure but can specifying overflow: hidden; to a container hide scroll bars?}'' seems to be related to UI rendering and was discarded during this step. 
This step discarded 6,235 comments and retained the remaining 1,337 comments for a detailed inspection.

 \emph{(Step III) Detailed Inspection:} In this step, two of the authors independently inspected the 1,337 review comments identified from the previous step, any subsequent discussion included in each review, and associated code contexts to determine whether a security defect was identified in each review. If any vulnerability is confirmed, the inspectors also classified it according to the CWE specification~\cite{cwe2020}. Similar to Bosu et al.~\cite{bosu-fse14}, we considered a code change vulnerable only if: (a) a reviewer indicated potential vulnerabilities, (b) our manual analysis of the associated code context found the code to be potentially vulnerable, and (c) the code author either explicitly acknowledged the presence of the vulnerability through a response (e.g., `\textit{Good catch}', `\textit{Oops!}') or implicitly acknowledged it by making the recommended changes in a subsequent patch.
Agreement between the two inspectors was computed using Cohen's Kappa ($\kappa$)~\cite{cohen1960coefficient}, which was measured as 0.94 (almost perfect\footnote{Cohen’s Kappa values are interpreted as following: 0 - 0.20 as
slight, 0.21 - 0.40 as fair, 0.41 - 0.60 as moderate, 0.61 - 0.80 as substantial, and 0.81 - 1 as almost perfect agreement}). Conflicting labels were resolved during a discussion session. At the end of this step, we identified total 516 code reviews that successfully identified security vulnerabilities.

\begin{table}
	\caption{Keywords to mine code reviews that identify security defect}
	\centering \label{table:security-keywords}
	\resizebox{\linewidth}{!} {
\begin{tabular}{|p{2.1cm}|p{1.8cm}|p{4.2cm}|}
\hline
\textbf{Vulnerability Type} & 
\textbf{CWE ID} & \textbf{Keywords*} \\
\hline
Race Condition  & 362 - 368 & race, racy \\
\hline
Buffer Overflow & 120 - 127 & buffer, overflow, stack, \textit{strcpy}, \textit{strcat}, \textit{strtok}, \textit{gets}, \textit{makepath}, \textit{splitpath}, \textit{heap}, \textit{strlen}\\
\hline
Integer\newline Overflow & 190, 191, 680& integer, overflow, signedness, widthness, underflow \\
\hline
Improper \newline Access & 22, 264, 269, 276, 281 -290 & improper, unauthenticated, gain access, permission, \textit{hijack}, \textit{authenticate}, \textit{privilege}, \textit{forensic}, \textit{hacker}, \textit{root} \\
\hline
Cross Site Scripting (XSS) & 79 - 87& cross site, CSS, XSS, \textit{malform},  \\
\hline
Denial of \newline Service (DoS) / Crash & 248, 400 - 406, 754, 755 &  denial service, DOS, DDOS, crash \\
\hline
Deadlock & 833 & deadlock\\
\hline
SQL Injection & 89 & SQL, SQLI, injection \\
\hline
Format String & 134& format, string, printf, scanf\\
\hline
Cross Site \newline Request Forgery & 352  & cross site, request forgery, CSRF, XSRF, forged \\ \hline
Encryption & 310, 311, 320-327 & \textit{encrypt}, \textit{decrypt}, \textit{password}, \textit{cipher}, \textit{trust}, \textit{checksum}, \textit{nonce}, \textit{salt}\\
\hline
Common \newline keywords & - & security, vulnerability, vulnerable, hole, exploit, attack, bypass, backdoor, threat, expose, breach, violate, fatal, blacklist, overrun, insecure, \textit{scare}, \textit{scary}, \textit{conflict}, \textit{trojan}, \textit{firewall}, \textit{spyware}, \textit{adware}, \textit{virus}, \textit{ransom}, \textit{malware}, \textit{malicious}, \textit{risk}, \textit{dangling}, \textit{unsafe}, \textit{leak}, \textit{steal} , \textit{worm}, \textit{phishing}, \textit{cve}, \textit{cwe}, \textit{collusion}, \textit{covert}, \textit{mitm}, \textit{sniffer}, \textit{quarantine}, \textit{scam}, \textit{spam}, \textit{spoof}, \textit{tamper}, \textit{zombie}    \\
\hline 
\multicolumn{3}{p{9cm}}{*Approximately half of the keywords in this list are adopted from the prior study of Bosu et al. \cite{bosu-fse14}. Keywords in \textit{italic} are our additions to this list.}

\end{tabular}
}
\end{table}

\subsection{Building a dataset of controls (i.e. vulnerabilities escaped during code reviews)}
We searched the Monorail-based bug tracking system hosted at: \url{https://bugs.chromium.org/},    to identify a list of security defects for the Chromium OS project. We used the bug tracker instead of the CVE database, since the bug tracker includes more detailed information for each security defect (e.g., link to fixing commit and link to Gerrit where the fix was code reviewed). Moreover, some of the security defects may not be reported in the CVE, if it was identified during testing prior to its public release.  We used the following five-step approach to build this dataset. We also illustrate this process using an example security defect: \#\textit{935175}.

 \emph{(Step I) Custom search:} We use a custom search (i.e., ({\tt Type=Bug-Security status:Fixed OS=Chrome}), to filter security defects for the Chromium OS projects with the status as `Fixed'. Our search result identified total 591 security defects. We exported the list of defects as a comma-separated values( i.e., csv) file, where each  issue is associated with a unique ID.

\begin{figure*}
	\centering  \includegraphics[width=0.70\linewidth]{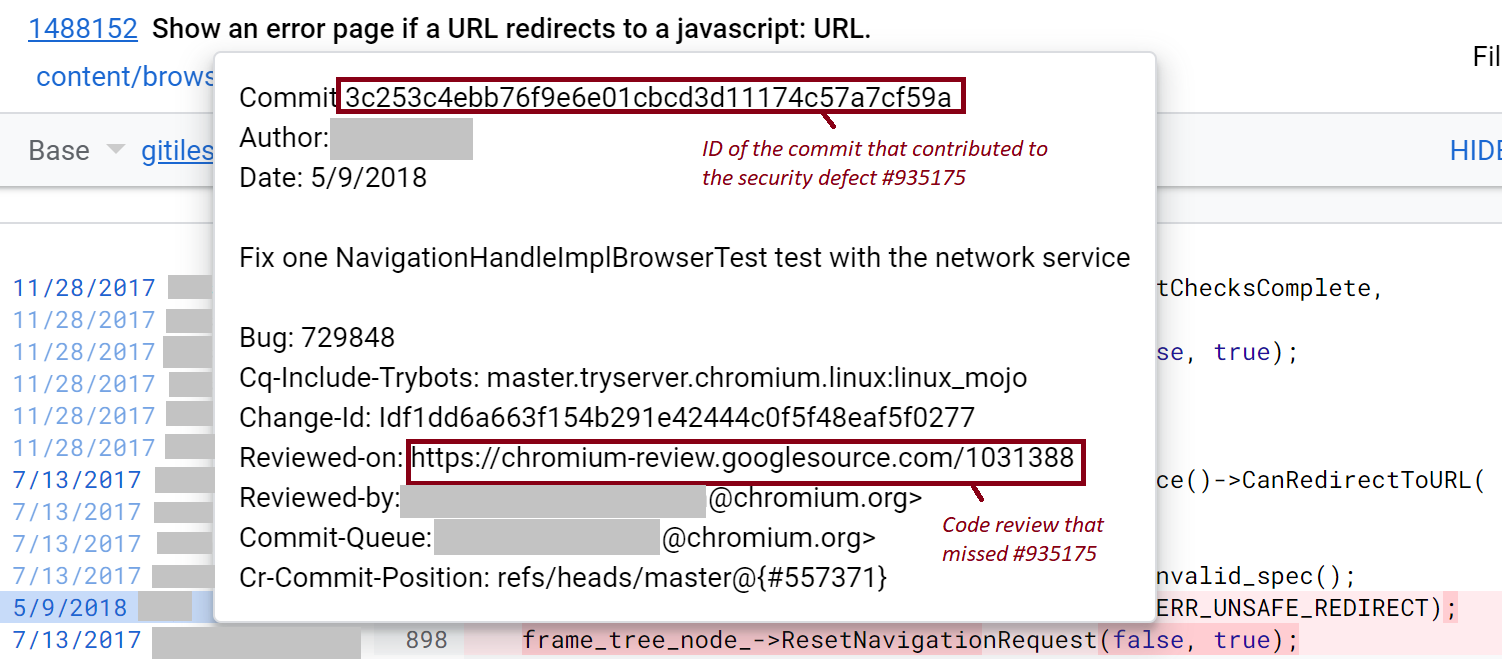}
	\caption{A vulnerability contributing commit (VCC) for the security defect \#\textit{935175} and the code review that missed it}
	\label{fig:vcc-commit}	
	\vspace{-12pt}
\end{figure*}

\begin{table*}
	\caption{Attributes of a code review that may influence identification of a vulnerability}
	\centering \label{table:attributes}
	\resizebox{\linewidth}{!} {
\begin{tabular}{|l|p{3cm}|p{6.5cm}|p{7cm}|}
\hline
\textbf{Type} &\textbf{Name} & \textbf{Definition} & \textbf{Rationale} \\
\hline

\multirow{3}{*}{Location}& Number of files under review & Number of files under review in a review request. & Changes that involve greater number of files are more likely to be defect-prone \cite{kononeko2015icsme}, yet more time-consuming to review.\\ \hhline{~---}
&Number of directory under review & Number of directory where files have been modified in a review request. & If the developers group multiple separate changes into a single commit, the review of those comprehensive changes could be harder. \\ \hhline{~---}
&Code churn & Number of lines added / modified / deleted in a code review. & Larger changes are more likely to have vulnerability \cite{bosu-fse14, nagappan-codechurn-icse05, nagappan2007using} and require more time to comprehend.\\ \hhline{~---}
&Lines of code & Numbers of lines of code in the file before fix. & Larger components are more  difficult to understand.\\
\hhline{~---}

&Complexity & McCabe's Cyclomatic Complexity \cite{mccabe1976complexity}. & Difficulty to comprehend a file increases with its cyclomatic complexity.\\ 
\hhline{~---}

&is\_bug\_fix & Code review request that is submitted to fix a bug & A bug fix review request may draw additional attention from the reviewers, as bugs often foreshadow vulnerabilities~\cite{camilo2015bugs}\\

\hline \hline
\multirow{3}{*}{Participant}&Author's coding experience & Number of code commits the author has submitted (i.e., both accepted and rejected) prior to this commit. & Experienced authors' code changes may be subject to less scrutiny and therefore may miss vulnerabilities during reviews. 
\\
\hhline{~---}
&Reviewer's reviewing experience & Number of code reviews that a developer has participated as a reviewer (i.e., code not committed by him/her) prior to this commit. & Experienced reviewers may be more likely to spot security concerns. \\
\hhline{~---}
&Reviewer's coding experience & Number of code commits that a reviewer has submitted (i.e., both accepted and rejected) prior to this commit. & Experienced developers provide more useful feedback during code review \cite{bosu2015characteristics} and may have more security knowledge.\\
\hline \hline

\multirow{4}{*}{Review process}

&Review time & The time from the beginning to the end of the review process.  We define the review process to be complete when the patchset is `Merged' to the main project branch or is `Abandoned'. & A cursory code review is more likely to miss security defects that require thorough reviews. \\ \hhline{~---}

& Number of reviewers involved $(NR_f)$ & Number of reviewers involved in reviewing file $f$ & As Linus's law suggest, the more eyeballs, the less likelihood of a defect remaining unnoticed. \\ \hhline{~---}



\hline\hline

\multirow{4}{*}{Historical}&Review ratio $(RR_{a,f})$ & The ratio between the number of prior reviews from developer $a$ to a file $f$ and the total number of prior reviews to that file. If the developer $a$ participated in $i$ of the $r$ prior reviews in file $f$ then: $RR_{a,f}$ = $\frac{i}{r}$ & A developer who has reviewed a particular file more may have better understanding of its design. \\
\hhline{~---}

&Commit ratio $(CR_{a,f})$ & The ratio between the number of commits to a file $f$ by author $a$ and the total number of commits to that file. If author $a$ makes $i$ of the $c$ prior commits then $CR_{a,f}$ = $\frac{i}{c}$ & A developer who makes frequent changes in a file may have better understanding of its design  \\ \hhline{~---}

&Weighted recent commits $(RC_{a,f})$ & If a file $f$ has total $n$ prior commits and author $a$ makes three of three of the prior n commits (e.g., $i, j, k$), where n denotes the latest commit, then: $RC_{a,f}$ = $\frac{(i+j+k)}{(1+2+3+...+n)}$ = $\frac{2(i+j+k)}{n(n+1)}$ & A developer who makes recent commits may have better understanding about the current design. \\ \hhline{~---}




&Total commit & Total number of commits made on the current file & Files that have too many prior commits might require extra attention from the reviewers. \\ \hhline{~---}


&Mutual reviews & Number of reviews performed by the current reviewer and author & Better understanding about the author's coding style might help the reviewer to investigate defects. \\ \hhline{~---}

&Number of review comments &Total number of review comments in the current file & Higher number of review comments indicate the file has gone through a more detailed review. \\ \hhline{~---}

&File ownership $(FO_{a,f})$ & The ratio between the number of lines modified by a developer and total number of lines in that file. If developer $a$ writes $i$ of total $n$ lines in file $f$, then $FO_{a,f}$ = $\frac{i}{n}$ & The owner of a file may be better suited to review that file. \\ 
\hline

\end{tabular}
}

\end{table*}

 \emph{(Step II)  Identifying vulnerability fixing commit:}
The Monorail page for each `Fixed' issue includes detailed information (e.g., commit\_id, owner, review URL,  list of modified files, and reviewer) regarding its fix. For example, \url{http://crbug.com/935175} details the information for the security defect \#\textit{935175} including the ID of the vulnerability fixing commit (i.e. `{ \tt 56b512399a5c2221ba4812f5170f3f8dc352cd74}'). We wrote a Python script to automate the extraction of the review URLs and commit\_ids for each security defect identified in Step I. Finally, we  excluded the security fixes that were not reviewed on Chromium OS's Gerrit repository (e.g., third-party libraries). At the end of this step, we were left with 239 security defects and its' corresponding fixes.

 \emph{(Step III)  Identifying vulnerability contributing commit(s):}
We adopted the modified version of the SZZ algorithm~\cite{Borg_2019} to identify the vulnerability introducing commits from the vulnerability fixing commits identified in Step II. Our modified SZZ algorithm uses the {\tt git blame } and {\tt git bisect } subcommands and is adopted based on the approaches followed in two prior studies\cite{perl2015vccfinder,meneely2013patch} on VCCs. For each line in a given file, the {\tt git blame } subcommand names the commit that last commit\_id that modified it. The heuristics behind our custom SZZ are as following:
\begin{enumerate}
    \item Ignore changes in documentations such as release notes or change logs.
    \item For each deleted / modified, blame the line that was deleted / modified, since if a fix needed to change a line, that often means that it was part of the vulnerability. 
    \item For every continuous block of code inserted in the bug fixing commit, blame the lines before and after the block, since security fixes are often done by adding extra checks, often right before an access or after a function call.
    \item If multiple commits are marked based on the above steps, mark commits as VCCs based on higher amount of lines until at least 80\% lines are accounted for.
\end{enumerate}

We manually inspect each of the VCCs identified by our modified SZZ algorithm as well as corresponding vulnerability fixing commits to exclude unrelated commits or include additional relevant commits. At the end of this step, we identified total 374 VCCs. Figure~\ref{fig:vcc-commit} shows a VCC for the security defect \#\textit{935175} identified through this process.

 \emph{(Step IV)  Identifying code reviews that approved VCCs:}
 A git  repository mirror for the Chromium OS project is hosted at \url{https://chromium.googlesource.com/} with a gitiles\footnote{https://gerrit.googlesource.com/gitiles/} based frontend. We used the REST API of gitiles to query this repository to download commit logs for each VCC identified in the previous step. Using a REGEX parser, we extract the URLs of the code review requests that approved VCCs identified in Step III. For example, Figure~\ref{fig:vcc-commit} also includes the URL of the code review that missed the security defect  \#\textit{935175}. At the end of this step, we identified total 374 code reviews that approved our list of VCCs.

 \emph{(Step V)  CWE classification of the VCCs:}
124 out of the 374 VCCs  in our dataset had a CVE reported in the NIST NVD database\footnote{\url{https://nvd.nist.gov/}} For example {\tt CVE-2019-5794} corresponds to the security defect \#\textit{935175}. For such VCCs, we obtained the CWE classification from the NVD database. For example, NVD classifies \#\textit{935175} as a `CWE-20: Improper Input Validation'. For the remaining 250 VCCs, two of the authors independently inspected each VCC as well as its fixing commits to  understand the coding mistake and classify it according to the CWE specification. Conflicting labels were resolved through discussions.

\subsection{Attribute Collection}
To answer the research questions motivating this study, we computed 18 attributes for each of the 890 code reviews (i.e. 516 cases + 374 controls). Majority of the attributes selected in this study have been also used in prior studies investigating the relationship between  software quality and code review attributes~\cite{mcintosh2014impact, thongtanunam2017review,kononeko2015icsme,krutauz2020code}. Table ~\ref{table:attributes} presents the list of our attributes with a brief description and rationale behind the inclusion of each attribute to investigate our research objectives. Those attributes are grouped into four categories: 1) vulnerability location, 2) participant characteristics, 3) review process, 4) historical measures.  We use several Python scripts and SQL queries to calculate those attributes from our curated dataset and our MySQL database of Chromium OS code reviews.

\section{Results}\label{sec:results}
Following subsections detail the results of the two research question introduced in the Section~\ref{sec:intro} based on our analyses of the collected dataset.

\subsection{RQ1: Which categories of security defects are more likely to be missed during code reviews?}
\label{sec:rq1-result}

For both identified and escaped security defects cases, we either obtained a CWE classification from the NVD database or manually assign one for those without any NVD reference. The 890 VCCs (i.e., both identified and escaped cases) in our dataset represented 86 categories of CWEs.
However, for the simplicity of our analysis, we decreased the number of distinct CWE categories by combining similar categories of CWEs into a higher level category. The CWE specification already  provides  a hierarchical categorization scheme\footnote{\url{https://cwe.mitre.org/data/graphs/1000.html}} to represent the relationship between different categories of weaknesses. For example, both CWE-190 (Integer Overflow or Wraparound) and CWE-468 (Incorrect Pointer Scaling) belong to the higher level category: CWE-682 (Incorrect Calculation). Using the higher level categories from the CWE specification~\cite{cwe2020}, we reduce the number of distinct CWE types in our dataset to 15. During this higher level classification,  we also ensured no common descendants among these final 15 categories.  Table ~\ref{table:cwe-distribution} shows the fifteen CWE categories represented in our dataset, their definitions, and both the number and ratios of identified /escaped cases,  in a descending order  based on their total number of appearances. 

The results of a Chi-Square ($\chi^2$) test suggest that some categories of CWEs are significantly ($\chi^2$=491.69, $p-value<0.001$) more likely to remain undetected during code reviews than the others.  
Chromium OS reviewers were the most efficient in identifying security defects due to `\textit{CWE-676: Use of potentially dangerous function}'. For example, following C functions are {\tt strcpy()}, {\tt strcat()}, {\tt strlen()}, {\tt strcmp()}, {\tt sprintf()} unsafe as they do not check for buffer length and may overwrite memory zone adjacent to the intended destination. As the identification of a CWE-676 is  relatively simple and does not require much understanding of the associated context, no occurrences of dangerous functions escaped code reviews. 
 Reviewers were also highly efficient in identifying security defects due to `\textit{CWE-404: Improper resource shut down or release}' that can lead to resource leakage. `\textit{CWE 682: Incorrect calculation}', which includes calculation of buffer size and  unsecured mathematical operation (i.e., large addition/multiplication or divide by zero), were also more likely to be identified during code reviews ($\approx$80\%).
 The other categories of CWEs that were more likely to be identified during code reviews include: improper exception handling (CWE-703) and synchronization mistakes (i.e., CWE- 662, and CWE-362).

On the other hand, Chromium OS reviewers were the least effective in identifying security defects due to insufficient verification of data authenticity (CWE-345), as all such occurrences remained undetected. Insufficient verification of data can lead to an application accepting invalid data.  Although Improper input validations (CWE-20) were frequent occurrences (i.e., 72), those remained undetected during $\approx$88\% code reviews. Improper input validation can lead to many critical problems such as uncontrolled memory allocation and SQL injection. Approximately 88\% security defects caused by improper access control (CWE-284) also remained undetected as reviewers were less effective in identifying security issues due to improper authorization and authentication, and improper user management.  The other categories of CWEs that were more likely to remain unidentified during code reviews include: operation on a resource after expiration or release (CWE-672) and exposure of resources to wrong spheres (CWE-668).

Our manual examinations of the characteristics of these CWE categories suggest that security defects that can be identified based on a few lines of the code context (e.g., unsafe method, calculation of buffer size, and resource release) are more likely to be identified during code reviews. On the other hand, Chromium OS reviewers were more likely to miss CWEs requiring either code execution (e.g., input validation) or understanding of  larger code contexts (e.g., resource lifetime, and authentication management ).

\vspace{-6pt}
 \begin{boxedtext}
\textbf{Finding 1:} \emph{The likelihood of a security defect's identification during code reviews  depends on its CWE category.}
\end{boxedtext}

\vspace{-6pt}
 \begin{boxedtext}
\textbf{Observation 1(A):} \emph{Security defects related to the synchronization of multi-threaded application, calculation of variable and buffer size, exception handling, resource release, and usage of prohibited functions were more likely to be identified during code reviews.}
\end{boxedtext}

\vspace{-6pt}
 \begin{boxedtext}
\textbf{Observation 1(B):} \emph{Security defects related to the user input neutralization, access control, authorization and authentication management, resource lifetime, information exposure, and datatype conversion were more likely to remain undetected during code reviews.}
\end{boxedtext}

\begin{table*}
	\caption{Distribution of Chromium OS CWEs identified /escaped code reviews}
	\centering \label{table:cwe-distribution}
	\resizebox{\linewidth}{!}{
\begin{tabular}{|r|p{9cm}|r|r|r|r|}
\hline
\textbf{CWE ID} & \textbf{CWE Definition} & \textbf{\#Identified} & \textbf{\%Identified} & \textbf{\#Escaped} & \textbf{\%Escaped} \\ \hline
662     &  Improper Synchronization      & 109                   & 89.34                & 13                 & 10.66             \\ \hline
362     &   Concurrent Execution using Shared Resource with Improper Synchronization ('Race Condition')     & 73                    & 87.95                & 10                 & 12.05             \\ \hline
682     &   Incorrect Calculation     & 65                    & 79.27                & 17                 & 20.73             \\ \hline
20      &    Improper Input Validation    & 9                     & 11.11                & 72                 & 88.89             \\ \hline
703     &   Improper Check or Handling of Exceptional Conditions     & 58                    & 72.50                & 22                 & 27.50             \\ \hline
404     &    Improper Resource Shutdown or Release    & 71                    & 94.67                & 4                  & 5.33             \\ \hline
284     &    Improper Access Control    & 8                     & 12.12                & 58                 & 87.88             \\ \hline
672     &    Operation on a Resource after Expiration or Release    & 3                     & 4.62                & 62                 & 95.38             \\ \hline
119     &    Improper Restriction of Operations within the Bounds of a Memory Buffer    & 42                    & 72.41                & 16                 & 27.59             \\ \hline
676     &    Use of Potentially Dangerous Function    & 53                    & 100.0                   & 0                  & 0.0                \\ \hline
668     &    Exposure of Resource to Wrong Sphere    & 15                    & 30.0                   & 35                 & 70.0                \\ \hline
704     &    Incorrect Type Conversion or Cast    & 1                     & 5.88                & 16                 & 94.12             \\ \hline
345     &    Insufficient Verification of Data Authenticity    & 0                     & 0.0                   & 13                 & 100.0                \\ \hline
665     &    Improper Initialization    & 5                     & 50.0                   & 5                  & 50.0                \\ \hline
19      &    Data Processing Errors    & 0                     & 0.0                   & 10                 & 100.0                \\ \hline
\end{tabular}
}
	\vspace{-8pt}
\end{table*}

\subsection{RQ2: Which factors influence the identification of security defects during a code review?}

To investigate this research question, we developed a logistic regression model. Logistic regression is very efficient in predicting a binary response variable based one or more explanatory variables \cite{bewick2005statistics}. In this study, we use the factors described in the  Table ~\ref{table:attributes} as our explanatory variables, while our response variable is a boolean that is set to \textit{TRUE}, if a code review's security defect was identified  by reviewer(s) and \textit{FALSE} otherwise. 

Being motivated by recent Software Engineering studies ~\cite{mcintosh2016empirical,thongtanunam2017review}, we adopt the model construction and analysis approach of Harrell Jr.~\cite{harrell2015regression}, which allows us to model nonlinear relationships between  the dependent  and explanatory variables more accurately. 
The following subsections describe our model construction and evaluation steps.

\subsubsection{Correlation and Redundancy Analysis}
\begin{figure}
	\centering  \includegraphics[width=\linewidth, trim={0 2cm 0 0},clip]{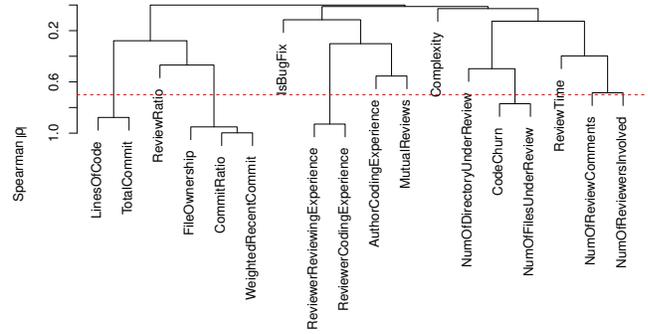}
	\caption{Hierarchical clustering of explanatory variables according to Spearman's $|\rho|$ and Sarle's VARCLUS. The dashed line indicates the high correlation coefficient threshold ($|\rho|=0.7$)}
	\label{fig:hierarchical-clustering}	
	\vspace{-12pt}
\end{figure}

If the explanatory variables to construct a model are highly correlated with each other, they can  generate a overfitted model. Following, Sarle's VARCLUS (Variable Clustering) procedure~\cite{sarle1990sas}, we use the Spearman's rank-order correlation test ($\rho$) \cite{statistics2013spearman} to determine highly correlated explanatory variables and construct a hierarchical representation of the variable clusters (Figure~\ref{fig:hierarchical-clustering}). We retain only one variable from each cluster of highly correlated explanatory variables. We use $|\rho \geq 0.7|$ as our threshold, since it has been recommended as  the threshold for high correlation~\cite{hinkle1998applied} and  has been  used as the threshold  in prior  SE studies~\cite{mcintosh2016empirical,thongtanunam2017review}.

We found four clusters of explanatory variables that have $|\rho| > 0.7$ -- (1) total lines of code (\textit{totalLOC}) and number of commits (\textit{totalCommit}), (2) commit ratio, weighted recent commit, and file ownership, (3) reviewer's coding experience and reviewer's reviewing experience, (4) amount of code churn, directory under review, and number of files under review. From the first cluster, we select total number of commit in the file. From the second cluster, we select file ownership. From the third and fourth cluster, we select reviewer's reviewing experience and number of directory under review respectively. 
Despite not being highly correlated, some explanatory variables can still be redundant. Since redundant variables can affect the modelled relationship between explanatory and response variables, we use the {\tt redun} function of the {\tt rms R} package with the threshold $R^2 \geq 0.9$ \cite{r-rms-package} to identify potential redundant factors among the remaining 12 variables and found none.

\subsubsection{Degrees of Freedom Allocation}
A model may be overfitted, if we allocate degrees of freedom more than a dataset can support (i.e., number of explanatory variables that the dataset can support). To minimize this risk, we estimate the budget for degrees of freedom allocation before fitting our model. As suggested by Harrell Jr. \cite{harrell2015regression}, we consider the budget for degrees of freedom to be $\frac{min(T,F)}{15}$, where T represents the number of rows in the dataset where the response variable is set to TRUE and F represents the number of rows in the dataset where the response variable is set to FALSE.  Using this formula, we compute our budget for degrees of freedom = 24, since our dataset has 516 TRUE instances and 374 FALSE instances. 

\begin{figure}
	\centering  \includegraphics[width=\linewidth, trim={0 0.5cm 0 0},clip]{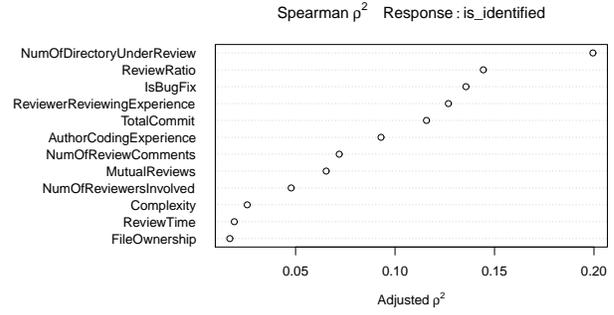}
	\caption{Dotplot of the Spearman multiple $\rho^2$ of each explanatory variable and the response. The larger values of $\rho^2$ indicate higher potential for a nonlinear relationship.}
	\label{fig:explanatory-vars}	
\end{figure}

For maximum effectiveness, we allocate this budget among all the survived explanatory variables in such a way that the variables that have more explanatory powers (i.e., explanatory variables that have more potential for sharing nonlinear relationship with the response variable) to be allocated with higher degrees of freedom than the explanatory variables that have less explanatory powers. To measure this potential, we  compute Spearman rank correlations ($\rho^2$) between the dependent variable and each of the 12 surviving explanatory variables (Figure ~\ref{fig:explanatory-vars}). Based on the results of this  analysis, we split the explanatory variables into two groups-- (1) we allocate three degrees of freedom to three variables, i.e., \textit{number of directory under review}, \textit{review ratio}, and \textit{reviewer's reviewing experience}, and (2) we allocate one degree of freedom for the remaining nine variables. Although \textit{isBugFix} has higher potential than \textit{reviewer's reviewing experience}, we cannot assign more than one degree of freedom for \textit{isBugFix} as it is dichotomous. As suggested by Harrell Jr. \cite{harrell2015regression}, we limit the maximum allocated degree of freedom for an explanatory variable below five to minimize the risk of overfitting.

\subsubsection{Logistic Regression Model Construction}
After eliminating highly correlated explanatory variables and allocating  appropriate degrees of freedom to the surviving explanatory variables, we fit a logistic regression model using our dataset. We use the {\tt rcs} function of the {\tt rms R} package \cite{r-rms-package} to fit the allocated degrees of freedom to the explanatory variables.

\subsubsection{Model Analysis}
After model construction,  we analyze the fitted model to identify the relationship between the response variable and each of the explanatory variables. We describe each step of our model analysis in the following.

\emph{Assessment of explanatory ability and model stability:} To assess the performance of our model, we use Area Under the Receiver Operating Characteristic (AUC) curve \cite{hanley1982meaning}. Our model achieves an AUC of 0.914. To estimate how well the model fits our dataset, we calculate Nagelkerke's Pseudo $R^2$ \cite{nagelkerke1991note}\footnote{ For Ordinary Least Square (OLS) regressions, $Adjusted\;R^2$ is used to measure a model's goodness of fit. Since it is difficult to compute $Adjusted\;R^2$  for a logistic regression model, the $Pseudo\;R^2$ is commonly used to measure its goodness of fit. The advantage of using Nagelkerke's $Pseudo\;R^2$ is that it's range is similar to the $Adjusted\;R^2$ range used for OLS regressions~\cite{smith2013comparison}.}. Our model achieves a $R^2$ value of $0.6375$, which is considered to be a good fit~\cite{nagelkerke1991note}.

\emph{Power of explanatory variables estimation:}
We use the Wald statistics (Wald $\chi^2$) to estimate the impact of each explanatory variable on the performance our model. We use the {\tt anova} function of the {\tt rms R} package to estimate the relative contribution (Wald $\chi^2$) and statistical significance ($p$) of each explanatory variable to the model. The larger the Wald $\chi^2$ value is, the more explanatory power the variable wields on our model. The results of our Wald $\chi^2$ tests (Table ~\ref{table:explanatory-power}) suggest that \textit{number of directory under review} wields the highest predictive power on the fitted model. \textit{ReviewRatio}, \textit{isBugFix}, \textit{ReviewerReviewingExperience}, and \textit{TotalCommit} are the next four most significant contributors. \textit{Number of review comments}, \textit{number of mutual reviews}, \textit{cyclomatic complexity of the file}, and \textit{review time} also wield significant explanatory powers. However, \textit{experience of code author}, \textit{number of reviewers involved in the review process}, and \textit{proportion of ownership of the file} do not  contribute significantly on the fitted model. 

We use the {\tt summary} function of the {\tt RMS R} package to analyze our model fit summary.   Table ~\ref{table:explanatory-power} also shows the contributions of each explanatory variable to fit our model using the  `\textit{deviance reduced by}' values.  For a generalized linear model,  deviance can be used to estimate goodness / badness of fit. A higher value of residual deviance indicates worse fit and a lower value indicates the opposite. A model with a perfect fit would have zero residual deviance. The {\tt NULL } deviance  value, which indicates how well the response variable is predicted by a model that includes only one intercept (i.e., the grand mean), is estimated as $1211.05$ for our dataset. The deviance of a fitted model decreases once we add explanatory variables. This decrement of residual deviance would higher for a variable with higher predictive power than for a variable with lower predictive power. For example, the explanatory variable ``\textit{directory\_under\_review}", which has the highest predictive power, reduces the residual deviance by 195.701 with a loss of three degrees of freedom. We can imply that the variable ``\textit{directory\_under\_review}" adds $\frac{195.701}{1211.05} \times 100\% = 16.16\%$ explanatory power to fit the model. Similarly, ``\textit{reviewers\_experience}" reduces the residual deviance by 104.104 with a loss of three degrees of freedom. Therefore, ``\textit{reviewers\_experience}" adds $\frac{104.104}{1211.05} \times 100\% = 8.6\%$ explanatory power to fit the model. Overall, our explanatory variables decrease the deviance by $571.73$ with a loss of 17 degrees of freedom. Hence, we can imply that our explanatory variables add $\frac{571.74}{1211.05} \times 100\% = 47.21\%$ explanatory power to fit the model which can be considered as a significant improvement over the null model.

\emph{Examination of variables in relation to response:}
Since Odds Ratio (OR) is recommended to measure the strength of relationship between an explanatory variable and the outcome~\cite{setia2016methodology}, we compute the OR of each explanatory variable in our fitted model (Table \ref{table:explanatory-power}) using 95\% confidence interval. In this study, the OR of an explanatory variable implies  how the probability of getting a true outcome (i.e., a vulnerability getting identified) increases with a unit change of that variable. Therefore, an explanatory variable with $OR > 1$ would increase the probability of a security defect getting identified during code reviews with its increment and vice versa. Since the explanatory variables used in our model have varying ranges (i.e., while `number of directory under review' varies between from 1 to 10, the `reviewing experience' varies between 0 to several hundreds), we cannot draw a generic conclusion by comparing the numeric OR value  of an explanatory variable against the OR of another variable that has a different range. 

Table \ref{table:explanatory-power} shows that the OR of `\textit{number of directory under review}' is 0.76 (i.e. $<1$), indicating that, if the `\textit{number of directory under review}' increases, a code review is more likely to miss a security defect. On the other hand, the odds ratio of the variables \textit{ReviewRatio} and \textit{IsBugFix} are well above $1$, which imply that if the review request is marked as a bug fix commit or the review conducts a significant number of prior review to that file, the security defect is more likely to be identified during code review. Since \textit{IsBugFix} is a dichotomous variable,  interpretation of its OR value (4.55) is straightforward. It indicates that vulnerabilities in a bug fix commit were 4.55 times more likely to be identified during code reviews than a non-bug fix commit. The results also suggest positive impact of review time on vulnerability identification.  Surprisingly, the overall reviewing experience of a developer does not increase his/her ability to identify security defects. 

\vspace{-6pt}
 \begin{boxedtext}
\textbf{Finding 2:} \emph{Our model has identified nine code review factors that  significantly differ between code reviews that successfully identified security defects and those  failed. Number of directories impacted by a code change has the most predictive power among those nine factors.}
\end{boxedtext}
 \begin{boxedtext}
\textbf{Observation 2(A):} \emph{The probability of a vulnerability getting identified during a code review decreases with the increase in number of directories, number of prior reviews by the reviewer,  number of prior commits in the file,  and number of review comments authored on a file during the current review cycle.}
\end{boxedtext}
 \begin{boxedtext}
\textbf{Observation 2(B):} \emph{The probability of a vulnerability getting identified during a code review increases with review time,  number of mutual reviews between the code author and a reviewer, cyclomatic complexity of the file under review, if the change belongs to a bug fix, and a reviewer's number of priors review with the file.}
\end{boxedtext}

\begin{table*}
	\caption{Explanatory powers of the code review attributes to predict the likelihood of a vulnerability to be identified during  code reviews}
	\centering \label{table:explanatory-power}
	\resizebox{\linewidth}{!} {\begin{tabular}{lrrrrrr}
\cline{2-7}
\multicolumn{1}{l|}{\textbf{}}                 & \multicolumn{1}{r|}{\textbf{Allocated D.F.}} & \multicolumn{1}{r|}{\textbf{Deviance}} & \multicolumn{1}{r|}{\textbf{Residual Deviance}} & \multicolumn{1}{r|}{\textbf{Deviance Reduced By (\%)}} & \multicolumn{1}{r|}{\textbf{Odds Ratio}}         & \multicolumn{1}{r|}{\textbf{Pr(\textgreater{}Chi)}} \\ \hline
\multicolumn{1}{|l|}{NULL}                     & \multicolumn{1}{r|}{}                        & \multicolumn{1}{r|}{}                  & \multicolumn{1}{r|}{1211.05}                    & \multicolumn{1}{r|}{}                                  & \multicolumn{1}{r|}{}                            & \multicolumn{1}{r|}{}                               \\ \hline
\multicolumn{1}{|l|}{NumOfDirectoryUnderReview} & \multicolumn{1}{r|}{3}                       & \multicolumn{1}{r|}{195.70}            & \multicolumn{1}{r|}{1015.35}                    & \multicolumn{1}{r|}{16.16}                             & \multicolumn{1}{r|}{{\color[HTML]{9A0000} 0.76}} & \multicolumn{1}{r|}{\textless 0.001***}             \\ \hline
\multicolumn{1}{|l|}{ReviewerReviewingExperience}    & \multicolumn{1}{r|}{3}                       & \multicolumn{1}{r|}{104.10}            & \multicolumn{1}{r|}{911.25}                     & \multicolumn{1}{r|}{8.60}                              & \multicolumn{1}{r|}{{\color[HTML]{9A0000} 0.99}} & \multicolumn{1}{r|}{\textless 0.001***}             \\ \hline
\multicolumn{1}{|l|}{ReviewRatio}            & \multicolumn{1}{r|}{3}                       & \multicolumn{1}{r|}{85.49}             & \multicolumn{1}{r|}{825.76}                     & \multicolumn{1}{r|}{7.06}                              & \multicolumn{1}{r|}{{\color[HTML]{036400} 1.83}} & \multicolumn{1}{r|}{\textless 0.001***}             \\ \hline
\multicolumn{1}{|l|}{IsBugFix}             & \multicolumn{1}{r|}{1}                       & \multicolumn{1}{r|}{67.14}             & \multicolumn{1}{r|}{758.62}                     & \multicolumn{1}{r|}{5.54}                              & \multicolumn{1}{r|}{{\color[HTML]{036400} 4.55}} & \multicolumn{1}{r|}{\textless 0.001***}             \\ \hline
\multicolumn{1}{|l|}{TotalCommit}            & \multicolumn{1}{r|}{1}                       & \multicolumn{1}{r|}{40.03}             & \multicolumn{1}{r|}{718.59}                     & \multicolumn{1}{r|}{3.30}                              & \multicolumn{1}{r|}{{\color[HTML]{9A0000} 0.97}} & \multicolumn{1}{r|}{\textless 0.001***}             \\ \hline
\multicolumn{1}{|l|}{NumOfReviewComments}    & \multicolumn{1}{r|}{1}                       & \multicolumn{1}{r|}{26.56}             & \multicolumn{1}{r|}{692.03}                     & \multicolumn{1}{r|}{2.19}                              & \multicolumn{1}{r|}{{\color[HTML]{9A0000} 0.98}} & \multicolumn{1}{r|}{\textless 0.001***}             \\ \hline
\multicolumn{1}{|l|}{ReviewTime}             & \multicolumn{1}{r|}{1}                       & \multicolumn{1}{r|}{23.86}             & \multicolumn{1}{r|}{668.17}                     & \multicolumn{1}{r|}{1.97}                              & \multicolumn{1}{r|}{{\color[HTML]{036400} 1.01}} & \multicolumn{1}{r|}{\textless 0.001***}             \\ \hline
\multicolumn{1}{|l|}{MutualReviews}          & \multicolumn{1}{r|}{1}                       & \multicolumn{1}{r|}{14.95}             & \multicolumn{1}{r|}{653.22}                     & \multicolumn{1}{r|}{1.23}                              & \multicolumn{1}{r|}{{\color[HTML]{036400} 1.01}} & \multicolumn{1}{r|}{\textless 0.001***}             \\ \hline
\multicolumn{1}{|l|}{Complexity}               & \multicolumn{1}{r|}{1}                       & \multicolumn{1}{r|}{12.17}             & \multicolumn{1}{r|}{641.05}                     & \multicolumn{1}{r|}{1.01}                              & \multicolumn{1}{r|}{{\color[HTML]{036400} 1.12}} & \multicolumn{1}{r|}{\textless{}0.001***}            \\ \hline
\multicolumn{1}{|l|}{AuthorCodingExperience}      & \multicolumn{1}{r|}{1}                       & \multicolumn{1}{r|}{0.95}              & \multicolumn{1}{r|}{640.10}                     & \multicolumn{1}{r|}{0.08}                              & \multicolumn{1}{r|}{0.99}                        & \multicolumn{1}{r|}{0.33}                           \\ \hline
\multicolumn{1}{|l|}{FileOwnership}          & \multicolumn{1}{r|}{1}                       & \multicolumn{1}{r|}{0.60}              & \multicolumn{1}{r|}{639.50}                     & \multicolumn{1}{r|}{0.05}                              & \multicolumn{1}{r|}{{\color[HTML]{000000} 1.52}} & \multicolumn{1}{r|}{0.44}                           \\ \hline
\multicolumn{1}{|l|}{NumOfReviewersInvolved} & \multicolumn{1}{r|}{1}                       & \multicolumn{1}{r|}{0.18}              & \multicolumn{1}{r|}{639.32}                     & \multicolumn{1}{r|}{0.02}                              & \multicolumn{1}{r|}{1.04}                        & \multicolumn{1}{r|}{0.67}                           \\ \hline
\textbf{Total}                                 & \textbf{18}                                  & \textbf{571.73}                        &                                                 & \textbf{47.21\%}                                         &                                                  &                                                    
\\
                                               \multicolumn{7}{l}{\begin{tabular}[c]{@{}l@{}}Statistical significance of explanatory power according to Wald $\chi^2$ likelihood ratio test:\\ * p \textless 0.05; ** p \textless 0.01; *** p \textless 0.001;\end{tabular}}

\end{tabular}
}
	\vspace{-8pt}
\end{table*}
\section{Implications}
\label{sec:implication}

In this section, we describe the implications of our findings.

\subsection{Findings from RQ1}
Table ~\ref{table:cwe-distribution} suggests that reviewers detect CWE-662 and CWE-362 in most of the cases. Both of the CWEs are related to the security issues for multi-threaded applications (Improper synchronization and race condition). Hence, we can infer that Chromium OS developers have adequate expertise in securing  multi-threaded programs. Developers are also detecting issues related to improper calculation of array buffer or variable size in most of the cases which can overcome the possibility of potential buffer and/or integer overflow/underflow. However, identifying security issues with user input sanitization remains a concern. Most of the issues related to improper input validation have been escaped during code review which can lead to security vulnerabilities such as SQL injection attack, cross-site scripting attach, and IDN holograph attack. Chromium OS project manager may tackle this problem in two possible ways. First, they may leverage additional quality assurance practices, such as static analysis, fuzzy testing that are known to be effective in identifying these categories of vulnerabilities. Second, education / training materials may be provided to reviewers to improve their knowledge regard these CWEs.

\subsection{Findings from RQ2}
Herzig and Zeller find that when developers commit loosely related code changes that affect all related modules, the likelihood of introducing bug increases \cite{herzig2013impact}. Such code changes are termed as \textit{tangled} code changes. Our study also finds that if the code change affects multiple directories, the security defect is more likely to escape code review. Reviewing tangled code changes can  be challenging due to difficulties in comprehension. To tackle this issue we recommend: 1) trying to avoid code changes dispersed across a large number of directories, when possible, 2) spending additional time during such changes as our results also suggest positive impact of review time on vulnerability identification, and 3) integrate a tool, such as the one proposed by Barnett et al.~\cite{barnett2015helping}  to help reviewers navigate tangled code changes.

Our results also suggest that Chromium OS reviewers, who have participated in higher number of code reviews for were less likely to identify security defects. There may be several possible explanations for this result. First, developers who participates in a large number of reviews may become less cautious (i.e., review fatigue) and miss security defects. Second,  developers who review lot of changes may have to spend less time per review, as code reviews are considered as secondary responsibilities in most projects. Therefore, such developers become less effective in identifying security defects due to hasty reviews.  Finally, identification of security defects may require special skillsets that do not increase a developer's participation in non-security code reviews.  
While we do not have a definite explanation, we would recommend project managers to be more aware of `review fatigue' and avoid overburdening a person with a larger number of reviews.

The likelihood of file's vulnerability escaping increases with the total number of commit it has encountered during its lifetime. A file with higher number of commits indicates more frequent changes in that file than others due to bugs or design changes. Since bugs often foreshadow vulnerabilities~\cite{camilo2015bugs}, developers should be more cautious while reviewing files that frequently go through modifications.

Interestingly, if a code change is marked as a bug fix, developers are more likely to identify security defects (if exists) during code reviews, which suggests  extra cautions during such reviews. Therefore, an automated model may  be used to predict and assign tags (e.g., `security critical') to code changes that are more likely to include vulnerabilities to draw reviewers' attentions and seek their cautiousness. 

Unsurprisingly, the likelihood of a security  defect getting identified increases with review time (i.e., time to conclude a  review). Although, taking too much time to complete a review would slow the development process, reviewers should make a trade-off between time and careful inspection, and try to avoid rushing reviews of security critical changes. 
The number of mutual reviews between a pair of developers also has a positive effect on the likelihood of security defect identification.  When two developers review each other's code over a period of time, they become more aware of each other's coding styles, expertise, strengths, and weaknesses. That awareness might  help one to pay attention to areas that he/she thinks the other has a weakness or where he/she may make a mistake. Since mutual reviews have positive impact, we recommend promoting such relationships. 
\section{Threats to Validity}
\label{sec:threats}

Since case control studies originate from the medical domain,one may question  whether we can use this study framework to study SE research questions. We would like to point out that prior SE studies have  adopted various research designs, such as systematic literature review, controlled experiment, ethnography,  and focus group  that have originated in other research domains.  Although, the results of this study do not rely on the case-control study framework, we decided to use this design, since: 1) our study satisfies the criteria for using this framework, and 2) following a established methodological framework strengthens an empirical research such as this study. 

Our keyword-based mining technique to identify whether a code review identifies security defect or not poses a threat to validity.  We may miss a security defect if the review comments do not contain any of the keywords that we used. However, as we are only considering those reviews that belong to security defects and ignoring the rest, we are considering that false-negative labeling of security defect will not make any impact on our study. Nevertheless, as we manually check all the security defects while assigning CWE ID, we find no false-positive labelling a code review as related to security defect. 

Another threat to is the categorization of CWE ID. As one of our authors manually checks all the codes to find weakness type and assign the best match CWE ID for each weakness, that author might categorize a weakness with a wrong or less suited CWE ID. To minimize the effect of this threat, another author randomly chooses 200 source code files and manually assign CWE ID following a similar process without knowing the previously labeled CWE ID. We find that 194 out of 200 labels fall in the same group of CWE IDs that were labeled earlier. So, we are considering that this threat will not make any significant change in our results. 

Another threat is the measure we take to calculate the developer’s experience. We can interpret the term ``experience'' in many ways. And in many ways, measuring of experience will be complex. For example, we cannot calculate the amount of contribution of a developer to other projects. Although a different experience measure may produce different results, we believe our interpretation of experience is reasonable as that reflects the amount of familiarity with current project. 

Finally, results based on a single project or even a handful of projects can be subject to lack of external validity. Given the manual work involved in the data collection process, it is often infeasible to include multiple projects. Moreover, historical evidence provides several examples of individual cases that contributed to discovery in physics, economics, and social science (see ``Five misunderstandings about case-study research'' by Flyvjerg~\cite{flyvbjerg2006five}). Even in the SE domain case studies of the Chromium project~\cite{camilo2015bugs,di2016security}, Apache case study by Mockus \textit{et al.}~\cite{mockus2000case}, and Mozilla case study by Khomh \textit{et al.}~\cite{khomh2012faster} have provided important insights. To promote building knowledge through families of experiments, as championed by Basili~\cite{basili1999building}, we have made our dataset and scripts publicly available~\cite{paul_rajshakhar_2021_4539891}.

\section{Related Work}
\label{sec:related-work}

Code review technologies are widely used in modern software engineering. Almost all the large scale projects have adopted peer code review practices with the goal of improving product quality \cite{rigby2013convergent}. Researchers have justified the benefit of code reviews to identify missed defects \cite{mantyla2008types, beller2014modern}. Prior studies also find that peer code review can be very effective in identifying security vulnerability \cite{bosu-fse14}. That is why developers use 10-15\% of their working hours in reviewing other's code \cite{bosu2016process}. However, despite the popularity and evidence in support, some researchers explore that peer code reviews are not always performed effectively, which decelerates the software development process \cite{czerwonka2015code}.

Despite putting lots of efforts in code review to keep product secured, a significant number of security vulnerability is reported every year and the number is ever-increasing. Although some prior studies \cite{edmundson2013empirical, beller2014modern} have questioned about the effectiveness of peer code review in identifying security vulnerabilities, they did not explore the factors that could be responsible for this ineffectiveness. Researchers have introduced several metrics of code reviews over time that can be used to identify security vulnerability \cite{meneely2014empirical,  meneely2013patch, alves2016software}. However, they did not investigate the state of those attributes when code review cannot identify security vulnerabilities. 

Meneely and Williams find that the engagement of too many developers to write a source code file can make that file more likely to be vulnerable; termed that situation as ``too many cooks in kitchen"  \cite{meneely2009secure}. But, they do not explore what characteristics of code review was responsible. Munaiah et al. use natural language processing to get the insights from code review that missed a vulnerability \cite{munaiah2017natural}. They investigate code review comments of Chromium project and find that code reviews that have discussions containing higher sentiment, lower inquisitiveness, and lower syntactical complexity are more likely to miss a vulnerability. To the best of our knowledge, no prior study has sought to identify the difference in security defects that are identified in code review and security defects that are escaped. Also, no prior studies introduce attributes that can be impactful in distinguishing code reviews where security defects get identified and code reviews where security defects get escaped.

\section{conclusion}\label{conclusion}
In this case-control study, we empirically build two datasets-- a dataset of 516 code reviews where security defects were successfully identified and a dataset of 374 code reviews where security defects were escaped. The results of our analysis suggest that the likelihood of a security defect's identification during code reviews  depends on its CWE category. A logistic regression model fitted on our dataset achieved an AUC score of 0.91 and has identified nine code review attributes that  influence identifications of security defects. While time to complete a review, the number of mutual reviews between two developers, and if the review is for a bug fix have  positive impacts on vulnerability identification, opposite effects are observed from the number of directories under review, the number of total reviews by a developer, and the total number of prior commits for the file under review. Based on the results of this study, we recommend: 1) adopting additional quality assurance mechanisms to identify security defects that are difficult to identify during code reviews, 2) trying to avoid tangled code changes when possible, 3) assisting the reviewers to comprehend tangled code changes, 4) balancing review loads to avoid review fatigue, and 4) promoting mutual reviewing relationship between developers.

\bibliographystyle{IEEEtranS}  
\bibliography{bibliography} 
\end{document}